\documentclass[11pt]{article}
\usepackage{mathtools}
\usepackage{hyperref}
\usepackage{graphicx}
\usepackage{graphics}
\usepackage{dsfont}
\usepackage{epsfig}
\usepackage{bbm}
\usepackage{amsmath,amssymb,amsthm,amscd}
\usepackage[sort&compress,square,numbers]{natbib} %%% This package was added
\usepackage{float}
\usepackage{braket}
\usepackage[arrow, matrix, curve]{xy}
\usepackage[latin1]{inputenc}
\usepackage{tikz}
\usepackage{mathabx}
\usetikzlibrary{shapes,arrows}
\usepackage{fancybox}
\usepackage{tikz}
\usetikzlibrary{matrix}
\usepackage{calc}
\usepackage{MnSymbol}
\usepackage[OT2,T1]{fontenc}
\DeclareSymbolFont{cyrletters}{OT2}{wncyr}{m}{n}
\DeclareMathSymbol{\Sha}{\mathalpha}{cyrletters}{"58}

\restylefloat{table}

\newcommand{\nn}{{\nonumber}}

\def\Mgn[#1]#2{{\overline{\cal M}_{#1,#2}}}
\def\pqs[#1,#2]{{\footnotesize{$\left[\begin{array}{c} #1\\#2  \end{array}\right]$}}} 
\def\pqsu[#1,#2]{\left[\begin{array}{c} #1\\#2  \end{array}\right]} 
\def\pqssu[#1,#2]{{\footnotesize{\left[\begin{array}{c} #1\\#2  \end{array}\right]}}} 
\def\pqh[#1,#2]{{\footnotesize{$\left[\begin{array}{c} #1\\#2  \end{array}\right]$}}} 
\def\pqhu[#1,#2]{\left[\begin{array}{c} #1\\#2  \end{array}\right]}

\newcommand{\ba}{\begin{eqnarray*}}
\newcommand{\ea}{\end{eqnarray*}}
\newcommand{\ban}{\begin{eqnarray}}
\newcommand{\ean}{\end{eqnarray}}
\newcommand{\be}{\begin{equation}}
\newcommand{\ee}{\end{equation}}
\newcommand{\ben}{\begin{equation}}
\newcommand{\een}{\end{equation}}
\numberwithin{equation}{section}

\numberwithin{equation}{section}

\begin{document}

\baselineskip=15pt

\begin{titlepage}
\begin{flushright}
\parbox[t]{1.73in}{\flushright 
UPR-1279-T}
\end{flushright}

\begin{center}

\vspace*{ 0.9cm}

{\large \bf \text{Discrete Symmetries in  Heterotic/F-theory Duality  and 
Mirror Symmetry 
%On Heterotic/F-Theory Duality with Abelian Gauge Symmetries
}}

\vspace*{ .2cm}
\vskip 0.5cm

\renewcommand{\thefootnote}{}
\begin{center}
 {Mirjam Cveti\v{c}$^{1,2,3}$,  Antonella Grassi$^{2}$, Maximilian Poretschkin$^1$}
\end{center}
\vskip .2cm
\renewcommand{\thefootnote}{\arabic{footnote}}

$\,^1$ {Department of Physics and Astronomy,\\
University of Pennsylvania, Philadelphia, PA 19104-6396, USA} \\[.3cm]

$\,^2$ {Department of Mathematics,\\
University of Pennsylvania, Philadelphia, PA 19104-6396, USA} \\[.3cm]
$\,^3$ {Center for Applied Mathematics and Theoretical Physics,\\ University of Maribor, Maribor, Slovenia}\\[.3cm]

{cvetic\ \textsf{at}\ cvetic.hep.upenn.edu,  grassi\ \textsf{at}\ math.upenn.edu,  mporet \textsf{at}\ sas.upenn.edu}

 \vspace*{0.0cm}

\end{center}

\vskip 0.0cm

\begin{center} {\bf ABSTRACT } \end{center}
We study aspects of Heterotic/F-theory duality for compactifications with Abelian discrete gauge 
symmetries.   We consider F-theory compactifications on genus-one fibered Calabi-Yau manifolds 
with   $n$-sections,  associated with the Tate-Shafarevich group $\mathbb{Z}_n$. Such models are obtained by studying first a specific toric set-up whose associated Heterotic vector bundle has structure group $\mathbb{Z}_n$. By employing a conjectured Heterotic/F-theory  mirror symmetry we construct dual geometries of these original toric models, where in the stable degeneration limit we obtain a discrete gauge symmetry of order two and three, for compactifications to six dimensions. We provide explicit constructions of mirror-pairs for symmetric examples with $\mathbb{Z}_2$ and $\mathbb{Z}_3$, in six dimensions. The Heterotic models with symmetric discrete symmetries are  related in field theory to a Higgsing of  Heterotic models with two symmetric abelian U(1) gauge factors, where  due to the St\"uckelberg mechanism only a diagonal U(1) factor remains massless, and thus  after Higgsing only a diagonal discrete symmetry of order $n$  is present in the Heterotic models and detected via Heterotic/F-theory duality. 
These constructions also provide further evidence for the conjectured mirror symmetry  in Heterotic/F-theory at the level of fibrations with torsional sections and those with multi-sections. 

\begin{flushright}
\parbox[t]{1.2in}{July, 2016}
\end{flushright}
\end{titlepage}

\tableofcontents

%\addtolength\topmargin{50pt}
%\addtolength\textheight{-105pt}

\section{Introduction and Summary of Results}

Recent years have witnessed important advances in F-theory compactifications\cite{Vafa:1996xn,Morrison:1996na, Morrison:1996pp}. While the study of non-Abelian gauge symmetries has been extensively studied in the past,  the study of Abelian and discrete gauge symmetries has been advanced only lately. 
 
 F-theory compactifications with Abelian gauge 
symmetries  U(1)$^n$ are based on constructions  of elliptically fibered Calabi-Yau 
manifolds with  rank-$n$ Mordell-Weil (MW) group of rational sections. A systematic approach was initiated in \cite{Morrison:2012ei} for constructions with  rank-one MW group, and generalized to  rank-two  \cite{Borchmann:2013jwa, Cvetic:2013nia, Cvetic:2015ioa} 
and rank-three \cite{Cvetic:2013qsa} MW groups, respectively.

Recently, there has also been progress in F-theory compactifications  with discrete gauge symmetries $\mathbb{Z}_n$, initiated  in \cite{Braun:2014oya} and further advanced in  \cite{Morrison:2014era, Anderson:2014yva, Klevers:2014bqa, Garcia-Etxebarria:2014qua,Cvetic:2015moa}.  A natural object  attached to these compactifications is given by the Tate-Shafarevich (TS) group of the genus-one fibration which is a discrete group that organizes inequivalent genus-one geometries which share the same associated Jacobian fibration. In F-theory discrete symmetries are also well  understood as a result of a Higgsing process of F-theory compactifications with U(1) symmetries, where the matter field with charge $n$ acquires a vacuum expectation value and breaks the abelian U(1) gauge symmetry to discrete $\mathbb{Z}_n$ one.  This field theoretical process is geometrically interpreted as a conifold transition \cite{Anderson:2014yva, Mayrhofer:2014haa, Mayrhofer:2014laa}.  Most of the past works  primarily  focused on $\mathbb{Z}_2$  gauge symmetry. However,  new insights into aspects of the TS group, and its relations to M-theory vacua, as well as the study of the Higgsing process in the case of $\mathbb{Z}_3$ were addressed in \cite{Cvetic:2015moa}. See also \cite{Anderson:2015cqy} for related work.

 Heterotic/F-theory duality plays an important role in shedding light on the origin of gauge symmetries in  Heterotic gauge theory from the geometric perspective of F-theory. In the past aspects of non-Abelian gauge symmetries have been studied extensively \cite{Vafa:1996xn,Morrison:1996pp,Morrison:1996na, Friedman:1997yq, Aspinwall:1998xj, Berglund:1998ej}.  In particular, Heterotic/F-theory duality allows for making statements about the Heterotic vector bundle $V$, which is typically hard 
to control, in terms of the controllable geometry of the Calabi-Yau manifold  on the 
F-theory side.   
On the other hand,  only recently key steps towards developing the geometrical duality map between Heterotic and 
F-theory compactifications with Abelian gauge symmetries were taken in  \cite{Cvetic:2015uwu}.

 The purpose of this paper is to study  discrete gauge symmetries in Heterotic/F-theory duality.  Since in Heterotic string theory the gauge group is given as the commutant of the structure group of the respective vector bundles within the two E$_8$ factors,  in order to engineer a discrete gauge group in a Heterotic string compactification, one needs a pair of vector bundles $V_1, V_2$ whose structure group is the centralizer of the desired discrete gauge group, e.g.,  \cite{Aspinwall:1998xj}.  

In order to construct explicitly such models we employ and generalize a conjecture with connects toric mirror symmetry and Heterotic/F-theory  mirror duality in eight dimensions \cite{Berglund:1998ej}. In particular, it relates the gauge symmetry (structure group of the bundle) of a K3 surface to the structure group of the Heterotic bundle (gauge symmetry) which is associated to the rational elliptic surfaces which arise from the dual K3 surface in the stable degeneration limit.  The dual pairs of K3 surfaces can be constructed by employing toric geometry techniques and are highlighted in Section 2.3. 
 For our purposes we construct the mirror models where the mirror symmetry in the fiber exchanges torsional sections of order-$n$  with the $n$-section. Thus, such a construction relates models with discrete structure group of the bundle to the model with a discrete gauge symmetry.

 We therefore start with the construction of toric models where in the stable degeneration limit the  Heterotic bundles contain a torsion of order $n$. By employing the  conjectured F-theory/Heterotic mirror-symmetry we construct dual toric models,   where in the stable degeneration limit we obtain a discrete gauge symmetry of order $n$, for compactifications in six dimensions. The explicit examples are based on the symmetric constructions where the two bundles $V_1$ and $V_2$  are the same.    In six dimensions for these models only a diagonal gauge symmetry is realized explicitly, and in particular for the dual polytope only a diagonal discrete symmetry is realized in the effective theory.
 We also demonstrate at the level of an effective six-dimensional field theory how such symmetric Heterotic models with discrete symmetry are related to un-Higgsing to Heterotic models with only a diagonal U(1) massless  gauge symmetry in the effective theory. 
 It is important to stress that this description therefore only works in complex codimensions greater than one.

 %The also elucidate from the field theory perspective that symmetric Heterotic models %with  discrete symmetries are  related via  Higgsing to the symmetric Heterotic models %with  abelian U(1) gauge factors. In the latter case   the St\''uckelberg mechanism for %U(1) factors ensures only a diagonal U(1) to remain massless, and thus after %Higgsing only a diagonal discrete symmetry of order $n$  is present in the Heterotic %models. It is this  diagonal discrete symmetry that is detected in the Heterotic/F-theory %duality in six-dimensions
 
 As concrete examples, we construct and analyse the mirror dual pairs for the case of  symmetric $\mathbb{Z}_2$ and $\mathbb{Z}_3$ symmetry. These constructions also provide further evidence for the conjectured mirror symmetry in Heterotic/F-theory at the level of fibrations with torsional sections and those with multi-sections.

This paper is organized in the following way. In Section 2, we provide a 
brief review of the key aspects  of Heterotic/F-theory duality, the origin of discrete symmetries in F-theory  and Heterotic theory, and a discussion of the mirror pairs of K3 surfaces, as it is  key in the study of conjectured mirror Heterotic/F-theory duality. In Section 3.1 we summarize a conjectured Heterotic mirror symmetry  where 
the Heterotic background bundle of the original K3 surface is interchanged with the 
gauge symmetry of the mirror K3 one.  In Section 3.2 we provide supporting evidence for this conjecture by studying mirror two-dimensional fiber ambient spaces where the torsional sections (associated with the discrete bundle structure group)  and multi-sections (associated with the discrete gauge symmetry) surfaces are interchanged. In Section 3.3  we elucidate in field theory aspects, how in six-dimensional Heterotic models with discrete symmetry are related  to those with U(1) gauge symmetry via Higgsing by matter fields  with conjectured charges;  in the symmetric case  only a diagonal U(1) is massless, and thus after Higgsing only a diagonal  discrete symmetry is present. In  Section 4.1 we present the construction of models with symmetric $\mathbb{Z}_2$ gauge symmetry which we demonstrate explicitly in the geometry of the mirror polytope.  In Section 4.2 we elaborate on aspects of six-dimensional geometry.  In Section 4.3  we repeat the construction for  models with the  symmetric  $\mathbb{Z}_3$ gauge symmetry. Concluding remarks in Section 5 highlight the key insights of the paper and possible future directions. In the Appendix we present explicit results for the Weierstrass map of the models  studied in the main text. 

\section{Heterotic/F-theory Duality and U(1)-Factors}
\label{sec:review+insights}

This section is divided into two parts:  In the first part, we review  basic facts about Heterotic/F-theory duality and discrete symmetries in F-theory and the Heterotic string. In addition, in order to set the stage for the construction of certain Heterotic background bundles, we also review the construction of mirror pairs of Calabi-Yau manifolds. 
The review part is mainly based on \cite{Aspinwall:1996mn, Friedman:1997yq, Andreas:1998zf}, 
to which we refer for further details. 

The next part formulates two important conjectures which are not rigorously proven but given strong evidence. The first conjecture is concerned with the construction of  background bundles  that have structure group (E$_7$ $\times$ SU(2))$/\mathbb{Z}_2$ and (E$_6$ $\times$ SU(3))$/\mathbb{Z}_3$. The second conjecture establishes a field theory connection between six-dimensional models with massive U(1) symmetries and models with discrete symmetries and geometrically corresponds to an analogue of conifold transitions on the Heterotic side.

%In this section, we present our strategy to derive Heterotic compactifications with background bundles of structure group E$_7$ $\times$ SU(2) and E$_6$ $\times$ SU(3). We start by recalling the basic statements of Heterotic/F-theory duality. Next, we recall the conjecture made in \cite{} that a mirror pair of K3 surfaces gives rise to two Heterotic compactifications where the gauge group and the structure group of the background bundle get exchanged.

\subsection{Heterotic/F-Theory duality in eight dimensions} 
\label{sec:hetF8d}

The basic statement of Heterotic/F-Theory duality is that the Heterotic E$_8\times$ E$_8$ String compactified on a torus, 
which we denote by $Z_1$, is equivalent to F-Theory compactified on an elliptically fibered K3 surface $X_2$. 

%Lower-dimensional dualities are obtained, applying the adiabatic argument 
%\cite{Vafa:1995gm}, by fibering the eight-dimensional duality over a 
%base manifold $B_{n-1}$ of complex dimension $n-1$ that is common to both theories of the 
%duality.

\subsubsection{The standard stable degeneration limit}   \label{stabledegeneration}

The moduli of both theories are matched in the stable degeneration limit. In this limit, the K3 surface $X_2$ degenerates into two half K3 surfaces $X_2^+$,
$X_2^-$ as
\be
	X_{2}\,\,\rightarrow\,\, X_{2}^{+}\cup_{Z_1}X_{2}^-\,.
\ee
$X_2^+$ as well as $X_2^-$ are elliptic fibrations $\pi_\pm: X_2^{\pm} \longrightarrow \mathbb{P}^1$ over 
a $\mathbb{P}^1$. These two $\mathbb{P}^1$ intersect in precisely one point so that the two 
half K3 surfaces intersect in a common elliptic fiber which is identified with the Heterotic elliptic 
curve, $X_2^+\cap X_2^-=Z_1$.

Traditionally, this has been formulated for K3 surfaces which are given as elliptic Weierstrass fibrations over a $\mathbb{P}^1$. More recently, it has been realized \cite{Cvetic:2015uwu, Anderson:2015cqy} that the study of U(1)s requires more general descriptions of the stable degeneration limit, in particular for fiber ambient spaces different from $\mathbb{P}^{(1,2,3)}$.

\subsubsection{Matching the continuous gauge groups} \label{matchingcontinuous}
~\\
\noindent 
The non-Abelian part of the F-theory gauge group is given by the singularities of the elliptic fibration of $X_2$, while the Abelian part is determined by the Mordell-Weil group of $X_2$ \cite{Vafa:1996xn,Morrison:1996na,Bershadsky:1996nh}.  These are inherited by the two half K3 surfaces $X_2^\pm$ in the stable degeneration limit as follows.

%The F-theory gauge group is given by the singularities of the elliptic fibration of $X_2$, 
%determining the non-Abelian part $G$, and its rational sections, which correspond to Abelian 
%gauge fields \cite{Vafa:1996xn,Morrison:1996na,Bershadsky:1996nh}. These are inherited by the two half K3 surfaces $X_2^\pm$ in the stable degeneration limit as follows. 

The homology lattice of a half K3 surface $X_2^\pm$ is given in 
general by
\be
H_2(X_2^\pm, \mathbb{Z}) = \Gamma_8 \oplus U \, .
\ee
Here, $\Gamma_8$ denotes the root lattice of E$_8$, while $U$ contains the classes of the elliptic fiber as well as of the zero section. For a non-generic half K3 surface, where the the curves in a sublattice $\mathcal{R}(G^\pm)$ of $\Gamma_8$, denoting the root lattice of some ADE group $G^\pm$, are shrunken to zero size, the half K3 surface develops a singularity of type $G^\pm$.   In rational homology, $\Gamma_8$ further splits as
\be
\Gamma_8 = \text{MW}(X_2^\pm) \oplus \mathcal{R}(G^\pm) \, ,
\ee
where MW$(X_2^\pm)$ denotes the Mordell-Weil group of $X_2^\pm$. 
%In addition, $\mathcal{R}(G^\pm)$ and $MW(X_2^\pm)$ sum up to the root lattice of E$_8$. 
%If the curves of $\mathcal{R}(G^\pm)$ are shrunken to zero size, the half K3 surface develops a singularity of type $G^\pm$. 

The latter fact provides the connection to the Heterotic description of the gauge group. According to \cite{Friedman:1997yq}, the moduli 
space of semi-stable E$_8$-bundles on an elliptic curve $Z_1$ corresponds to the complex 
structure moduli space of a half K3 surface $X_2$ whose anti-canonical class is given by 
$Z_1$. Furthermore, if $X_2$ has an ADE singularity of type $\tilde{G}_\pm$ then the structure 
group of $V_1$, $V_2$ is reduced to the centralizer $H_\pm$ of $\tilde{G}_\pm$ within 
E$_8$, respectively. 
In contrast, the abelian U(1) symmetries are translated as follows. Sections of MW$(X_2^+)$ that glue with a section of MW$(X_2^-)$ give rise to a global section of the K3 surface $X_2$ \cite{Cvetic:2015uwu}. 

The six-dimensional duality is obtained by fibering the eight-dimensional duality over a common $\mathbb{P}^1$. Thus, on the F-theory side one deals with a Calabi-Yau threefold $Z_3$ which is elliptically fibered\footnote{This is the classical approach. More generally, one is led to consider genus-one fibered three-folds as well.} over a Hirzebruch surface $\mathbb{F}_n$. In contrast, the Heterotic string is compactified on an elliptically fibered K3 surface whose base $\mathbb{P}^1$ is to be identified with the base of the Hierzebruch surface $\mathbb{F}_n$. In general, due to monodromies on the base also non-simply laced gauge groups can occur\cite{Friedman:1997yq, Bershadsky:1996nh}. In addition, only those eight-dimensional sections that promote to rational six-dimensional sections give rise to U(1) symmetries in six dimensions\footnote{In general, if a half K3 surface has a singularity of rank $r$, there are $9-r$ linear independent sections in its Mordell-Weil group. However, once the half K3 surface is promoted to a six-dimensional rational three-fold, these sections are not necessarily preserved.}.

Finally, another six-dimensional effect is the non-perturbative enhancement of the gauge group \cite{Bershadsky:1996nh} due to singularities which are not localized within the fiber $\mathbb{P}^1$ of $\mathbb{F}_n$. In other words, these are singularities which are visible both on the Heterotic side as well as on the F-theory side.

\subsection{Discrete Symmetries in Heterotic String Theory and F-theory}

As the main focus of this work is the investigation of discrete symmetries, we review their appearance on the F-theory side as well as on the Heterotic side in this subsection.

\subsubsection{Discrete symmetries in F-theory} \label{discretesymmetriesinFtheory}
Discrete symmetries are best understood within F-theory as result of a Higgsing process of continuous U(1) symmetries. It is important to stress that this description therefore only works in complex co-dimensions greater than one. As a well-known fact, U(1) symmetries within F-theory are detected by the generators of the Mordell Weil group \cite{Morrison:2012ei}. Each such generator gives rise to an additional globally well-defined embedding of the base manifold into the elliptic fibration. Reducing the $C_3$-form field along the corresponding divisors identifies the corresponding U(1) symmetry. 

Such a U(1) symmetry can be higgsed to a discrete $\mathbb{Z}_n$ symmetry, if there is a matter field of charge $n$. Such matter fields arise from M2-branes that wrap components of  $I_2$-fibers which appear at co-dimension two loci. Their corresponding charges are determined from the number of intersections of the corresponding sections with the respective component of the $I_2$-fiber. These fields become massless if the corresponding component of the $I_2$-fiber shrinks to zero size. From a mathematical perspective, this shrinking can be viewed as part of a conifold transition \cite{Anderson:2014yva, Mayrhofer:2014haa, Mayrhofer:2014laa}. Here, the shrunken component of the $I_2$-fiber, which is topologically a two-sphere, gets replaced by a three-sphere. The physical interpretation of the latter deformation is to give a vacuum expectation value to the Higgs field. In addition, for an $\mathbb{Z}_n$-symmetry, the conifold transition glues $n$ rational sections to an $n$-section. 
In fact, genus-one fibered Calabi-Yau manifolds with an $n$-section have an element of order $n$ in their Tate-Shafarevich (TS) group which is the geometrical analogue of the $\mathbb{Z}_n$-symmetry that occurs in the fields theory and labels in-equivalent geometries that share the same Jacobian fibration \cite{Braun:2014oya, Morrison:2014era, Cvetic:2015moa}.

\subsubsection{Discrete symmetries in the Heterotic string} \label{subsectiondiscretesymmetries}

As discussed in a previous subsection \ref{matchingcontinuous}, the Heterotic gauge group is given as the commutant of the structure group of the respective vector bundles within the two E$_8$ factors. Thus, in order to engineer a discrete gauge group in a Heterotic string compactification, one needs a pair of vector bundles $V_1, V_2$ whose structure group is the centralizer of the desired discrete gauge group. These centralizers have been determined for different discrete groups in, e.g., \cite{Aspinwall:1998xj}. E.g., in order to realize a discrete gauge group $\mathbb{Z}_2$, one needs a background bundle with structure group $($E$_7$ $\times$ SU(2)$)/\mathbb{Z}_2$. 
%A Heterotic string compactification requires, among other parameters, the choice of two semi-stable background vector bundles $V_1, V_2$ that carry generically the structure group E$_8$ $\times$ E$_8$. In an eight-dimensional compactification, the Heterotic string is compactified on a torus $E$. As observed in \cite{}, the moduli space of semi-stable vector bundles with structure group E$_8$ is given by the complex structure moduli space of a half K3 surface $S$ whose anti-canonical class is given by $E$. If $S$ happens to have an ADE-singularity of type $G$, then the corresponding structure group is reduced to the commutant of $G$ within E$_8$. In conclusion, the physical gauge group is given by the singularities of the two half K3 surfaces $S_1, S_2$ which correspond to the two vector bundles $V_1, V_2$. 

\subsection{Constructing mirror pairs of K3 surfaces}

In the following, we recall the construction of toric mirror pairs of K3 surfaces. As outlined in section \ref{backgroundbundlesmirrorsymmetry}, we will eventually use mirror symmetry techniques in order to construct background bundles with structure groups (E$_7$ $\times$ SU(2)$)/\mathbb{Z}_2$ and (E$_6$ $\times$ SU(3)$)/\mathbb{Z}_3$. To set the geometrical stage, we review in the following Batyrev's formalism \cite{Batyrev:1994hm} to construct mirror pairs of Calabi-Yau manifolds using pairs of reflexive polyhedra. A more detailed review on this subject and further references can be found in, e.g., \cite{Hosono:1994av}.

It is a well-known fact (see, e.g., \cite{cox1999mirror}), that given a reflexive $n$-dimensional polyhedron $\Delta$, there is a natural simplicial fan associated to it which will be denoted by $\Sigma$. $\Sigma$ defines a toric variety which is denoted by $\mathbb{P}_{\Delta}$. In particular, if a fine triangulation of $\Delta$ has been chosen, the associated variety $\mathbb{P}_{\Delta}$ is Gorenstein and terminal. We also note that a general section $\chi$ of the anti-canonical bundle $\mathcal{O}\left(-K_{P_{\Sigma}} \right)$ of $\mathbb{P}_{\Delta}$ defines a Calabi-Yau $(n-1)$-fold. Finally, there is a mirror Calabi-Yau $(n-1)$-fold which is given by a section of the anticanonical bundle of the toric variety associated to the dual polytope of $\Delta$, denoted by $\Delta^\circ$. In particular, the defining equation for $\chi$ is given by
\be
\chi = \sum_{P^\circ \in \Delta^\circ} \prod_{P \in \Delta} a_P x_{P}^{\langle P, P^\circ \rangle + 1} \, . \label{CYformula}
\ee
Here, $P$ and $P^\circ$ label the integer points of $\Delta$ and $\Delta^\circ$, respectively. In addition, $x_P$ denotes the coordinate associated to the ray determined by the point $P$ and $\langle, \rangle$ denotes the natural product between the dual lattices into which $\Delta$ and $\Delta^\circ$ are embedded. 
In addition, one can calculate the rank of the Picard lattice from Batyrev's formula which has been generalized to toric K3 surfaces in \cite{2010arXiv1011.1003B}
\begin{eqnarray}
h^{(1,1)}(X) &=& l(\Delta) -n - 1 - \sum_{\Gamma} l^*(\Gamma) + \sum_{\Theta} l^*(\Theta)l^*({\hat \Theta}) \, . \label{homologydimension}
%h^{(n-2,1)}(X) &=& l(\Delta) -n - 1 - \sum_{\Gamma} l^*(\Gamma) + \sum_{\Theta} l^*(\Theta)l^*({\hat \Theta}) \, .
\end{eqnarray}
Here $l(\Delta)$  ($l^*(\Delta)$) denote the number of (inner) points of the $n$-dimensional polytope $\Delta$. In addition, $\Gamma$ ($\Gamma^\circ$) denote the codimension one faces of $\Delta$ ($\Delta^\circ$), while $\Theta$ denotes a codimension two face with $\hat \Theta$ being its dual. 
In the following, we focus on K3 surfaces which are given as elliptic and genus-one fibrations over $\mathbb{P}^1$ and whose corresponding ambient space is given by the direct product $\mathbb{P}^1 \times \mathbb{P}_{\Delta_2}$, where $\Delta_2$ denotes any two-dimensional reflexive polytope. 
%The following discussion focuses on K3 surface which are constructed as genus-one fibrations over the base manifold $\mathbb{P}^1$. From the point of view of toric geometry, one is led to the investigation of toric fibrations of the fiber ambient space $\mathbb{P}_{\Sigma_2}$. There are 16 possible candidates which serve as an ambient space for the fibration, but only a few of them leads to a genus-one fibration. 

%\subsection{Heterotic/F-theory duality}

%The central statement of Heterotic/F-theory duality is the equivalance of the Heterotic string compactified on a torus $E$ with an F-theory compfactification on a K3 surface $X_2$. The precise translation of the data of the two theories is manifest in the so-called stable degeneration limit of the K3 surface $X_2$. At this point of the moduli space, the K3 surfaces splits into two half K3 surfaces, $X_2^+$ and $X_2^-$ that intersect in the common genus one fiber $E$ which is to be identified with the torus on which the Heterotic string has been compactified. In addition, the singularities of the two half K3 surfaces precisely determine the gauge group that originates from the breaking of the two E$_8$ factors. 

%As discussed in subsection \ref{subsectiondiscretesymmetries}, the 
\section{Conjectures in Field Theory and Geometry}

This section is devoted to the discussion of two conjectures. The first one is concerned with the construction of Heterotic vector bundles that exhibit structure groups whose commutant within E$_8$ gives rise to a discrete symmetry. The second one discusses the relation of six-dimensional Heterotic field theories with U(1)'s to those with discrete symmetries. Both conjectures are supported by a number of convincing observations.

\subsection{Constructing background bundles using mirror symmetry} \label{backgroundbundlesmirrorsymmetry}

In \cite{Berglund:1998ej}, explicit descriptions of vector bundles with structure groups of type ABCDE have been provided. In particular, Berglund and Mayr are considering K3 surfaces which are given as elliptic fibrations over $\mathbb{P}^1$, where the elliptic fiber is specified by the ambient space $\mathbb{P}^{(1,2,3)}$. Calling the affine base coordinate $z$, they conjecture the following statement. If a K3 surface has singularities of type $G_1, G_2$ at $z=0$ and $z=\infty$, respectively, its mirror K3 will have singularities of type $H_1$ and $H_2$, where $H_i = [\text{E}_8, G_i]$. 

This statement can be traced back to the fact that mirror symmetry for K3 surfaces has an interpretation in terms of orthogonal lattices. Considering the stable degeneration limit as discussed in section \ref{stabledegeneration} and further elaborated on in \cite{Cvetic:2015uwu}, it is clear that the points $z=0$ and $z= \infty$ map to different half K3 surfaces. Thus, each half K3 surface inherits precisely one singularity, whose commutant within E$_8$ gives rise to a singularity within the mirror dual K3 surface. In general, this conjecture should hold true more generally, i.e. in particular also for singularities that are inherited by both half K3 surfaces, as follows. 

Consider a K3 surface $X_2$ that has a couple of ADE singularities whose product is called $G$. The product of those singularities that are inherited by the first half K3 surface $X_2^+$ will be called $G_1$, and the definition of $G_2$ analogously applies to the second half K3 surface $X_2^-$. The mirror K3 surface $\tilde X_2$ will degenerate into two half K3 surfaces $\tilde X_2^+ \cup \tilde X_2^-$. The singularity content of $\tilde X_2^+$ and $\tilde X_2^-$ is called $H_1$ and $H_2$, respectively and will be given as $H_i = [\text{E}_8, G_i]$. It is important to stress that for this generalization, one should admit any toric ambient space $\mathbb{P}_{\Delta_2}$ - in contrast to only considering $\mathbb{P}^{(1,2,3)}$ - for a two-dimensional polytope $\Delta_2$ for the corresponding genus-one fibration. In this way, one also obtains structure groups which do not appear in the classification by Berglund and Mayr.

\subsection{Mirror symmetry in the fiber: Trading multi- for torsional sections}

In this sub-section we provide supporting evidence for the conjecture made above. In particular, as demonstrated in the concrete examples in section \ref{examples}, mirror symmetry exchanges different two-dimensional fiber ambient spaces, which leads to the exchange of multi-sections and torsional sections in the corresponding mirror geometries. 

In section \ref{examples} we will use this conjecture in order to construct background bundles with structure group $($E$_7$ $\times$ SU(2)$)/\mathbb{Z}_2$ and $($E$_6$ $\times$ SU(3)$)/\mathbb{Z}_3$. In particular, we make the following supporting observations. Starting with a K3 surface that has gauge group $(($E$_7$ $\times$ SU(2)$)/\mathbb{Z}_2)^2$, the ambient space of the elliptic fiber is required to be $\mathbb{P}^{(1,1,2)}/\mathbb{Z}_2$. The fiber ambient space of the dual K3 surface is given by $\mathbb{P}^{(1,1,2)}$ which generically leads to a genus one fibration that exhibits a bi-section. The latter is a clear signal of a $\mathbb{Z}_2$ symmetry. The same observation can be made for the case of a discrete $\mathbb{Z}_3$ symmetry where the fiber ambient spaces $\mathbb{P}^2/\mathbb{Z}_3$ and $\mathbb{P}^2$ get exchanged under mirror symmetry. 

However, as previously explained  in \ref{discretesymmetriesinFtheory}, discrete symmetries are best understood from a Higgsing perspective in six dimensions. Thus, it is expected that the corresponding multi-sections become apparent in six dimensions. This is indeed, what we will observe.

We would also like to stress that our approach sheds some light on the following observation made in  \cite{Klevers:2014bqa} and further confirmed in
  \cite{Braun:2014qka, Oehlmann:2016wsb}. That reference has constructed for every two-dimensional reflexive polytope a corresponding Calabi-Yau three-fold which was realized as an elliptic/genus-one fibration, specified by the corresponding two-dimensional polytope, over an arbitrary two-dimensional base. It was furthermore observed that mirror symmetry exchanges fiber ambient spaces that give rise to Calabi-Yau threefolds with non-trivial Tate-Shaverevich groups with fiber ambient spaces that give rise to a three-fold that exhibits non-trivial torsional sections.

Using the Heterotic/F-theory duality, one can explain this phenomenon as follows. It has been shown in \cite{Aspinwall:1998xj} that the construction of a discrete gauge symmetry $D$ in the Heterotic string requires background bundles with structure group of the form of $G/D$ where $G$ is a group of ADE type. Groups of the type $G/D$ require that the Mordell Weil group of the corresponding F-theory compactification has the torsional subgroup $D$. Due to the extended conjecture by Berglund and Mayr, the mirror Calabi-Yau manifold should exhibit a discrete symmetry of type $D$, where $D$ is a subgroup of the Tate-Shafarevich group of the mirror Calabi-Yau manifold.
\subsection{Heterotic field theory perspective}

In this section we discuss the conjectured field theory perspective on the Heterotic side which relates the models with U(1) factors to those with discrete gauge symmetries.  The approach is conjectured as we  did not calculate the charges of the matter fields, responsible for Higgsing of  the model with an  Abelian gauge symmetry to a discrete  one.
In particular, we shall focus on the examples where  the two U(1) factors appear symmetrically. i.e. in our approach we consider the U(1) which arise upon commutation of U(1) background bundles that are embedded symmetrically into the two respective E$_8$-bundles. 

%\\

\subsubsection {St\"uckelberg Mechanism} \label{Stuckelbergmechanism}
In the Heterotic string theory in 
in six and lower dimensions,  a geometric { St\"uckelberg mechanism}  can render a 
U(1) gauge field massive \cite{Green:2012pqa}.  In six dimensions the mass term of  U(1),  is due to  the modified ten-dimensional kinetic term of the Kalb-Ramond 
field $B_2$,   which upon dimensional  reduction  on an elliptically fibered Calabi-Yau threefold $Z_3$ and a U(1) background bundle, results in a six-dimensional  
kinetic term for the axions $\rho_\alpha$:
\be 
\mathcal{L}_{\text{St\"uck.}} = G^{\alpha\beta}\left(d\rho_\alpha + k_\alpha A_{U(1)} \right)\wedge \star\left(d\rho_\beta + k_\beta A_{U(1)} \right)\, ,\label{stuck}
\ee
where 
\be
	G^{\alpha\beta}=\int_{Z_3}\omega^\alpha\wedge \star \omega^\beta\,.
\ee
Here  $\omega^\alpha$, $\alpha=1,\ldots, b_2(Z_3)$, is a basis
of harmonic two-forms in $H^{(2)}(Z_3)$, where $b_2(Z_3)$ is the second Betti number of 
$Z_3$, the axions $\rho_\alpha$ are associated with the expansion of the Kalb-Ramond field $B_2$.  and the   $k_\alpha$ are flux quanta associated with the expansion of the U(1) bundle background field strength  $\mathcal{F}$ . Note, 
$\mathcal{F}=\frac{1}{2\pi i}c_1(\mathcal{L})$  where $c_1(\mathcal{L})$  is the first 
Chern class  the corresponding U(1) line bundle $\mathcal{L}$.

From \eqref{stuck}  a single U(1) gauge field will be  in general massive massive if we have a  non-trivial  first Chern class  of the U(1) bundle. However, in the presence of multiple massive 
U(1) gauge fields appropriate linear combinations of them, which belong to the kernel of the massive matrix, can 
remain massless U(1) fields.\footnote{For further details, see, e.g., \cite{Cvetic:2015uwu}. For similar computations, see,
e.g., ~\cite{Blumenhagen:2005ga}, where also the case of multiple U(1)'s is systematically discussed.} 

In the case of the symmetric example with  two U(1) factors, U(1)$_1$ and U(1)$_2$,  it is evident that a gauge boson associated with a symmetric linear combination, U(1)$_1 + $U(1)$_2$, remains massless, while the orthogonal  one, associated with the anti-symmetric  linear combination, U(1)$_1 -$U(1)$_2$,  becomes massive.   Thus, even though the geometry under stable degeneration indicates two U(1) factors in Heterotic theory, the St\"uckelberg mechanism ensures that only a symmetric  combination  of two U(1) gauge fields remains massless. It is this latter one, which is identified in the Heterotic/F-theory duality.

\subsubsection{Higgsing of symmetric U(1) Model} \label{HiggsingofU1}

The symmetric  Heterotic model with $\mathbb{Z}_Q$ discrete symmetry  can be  obtained from the one with symmetric U(1) gauge symmetry factors, by Higgsing  the latter model with the matter fields, $\phi_1$ and $\phi_2$.     Due to the symmetry structure of the model we conjecture that the $\phi_1$ and $\phi_2$ have integer charge $Q$ under  charged under  U(1)$_1$ and U(1)$_2$, respectively: 
\begin{center}
\begin{tabular}{|c|c|c|}\hline
& U(1) & U(1) \\\hline
$\phi_1$ & Q & 0 \\\hline
$\phi_2$ & 0 & Q \\\hline
\end{tabular}
\end{center}
After having acquired  non-zero vacuum expectation values, these field break the symmetry down to $\mathbb{Z}_Q\times \mathbb{Z}_Q$. 
%E.g.,  in the case of $\mathbb{Z}_2\times \mathbb{Z}_Q$ symmetry, $Q=2$. 

In particular, the two Higgs fields can be represented as
\be
\phi_1 = \left(V_1+\beta_1\right) e^{i \alpha_1 g}\, , \qquad \phi_2 = \left(V_2+\beta_2\right) e^{i \alpha_2 g}\, .
\ee
where $V_1$ and $V_2$ are the respective vacuum expectation values,    $\alpha_1$ and $\alpha_2$ are real fields and are identified with the Goldstone axions, while   $\beta_1$ and $\beta_2$  are the  Higgs fields.

In this representation the covariant kinetic energy terms for the matter fields  $\phi_1$ and $\phi_2$ take the form:
\be
\mathcal{L}_1 = \sum_{i=1}^2 | V_i+\beta_i|^2 g^2 \left(\partial_\mu \alpha_i - Q A_i \right)^2 + \sum_{i=1}^2 \left(\partial_\mu \beta_i\right)^2 \, .  \label{covariant}
\ee
Note that this Lagrangian  has a manifest discrete gauge symmetry $\mathbb{Z}_{Q}\times \mathbb{Z}_{Q}$. 

However,  due to the St\"uckelberg mechamism as discussed in \ref{Stuckelbergmechanism}, the  anti-symmetric combination of the gauge fields $A_- \equiv \frac{1}{\sqrt 2}\left( A_1-A_2\right)$  acquires the mass.  The  contribution   (\ref{stuck}) to the effective Lagrangian  can schematically written as:
\be
\mathcal{L}_2 = M^2 \left(\partial_\mu b - A_- \right)^2 \, , \label{stuck2}
\ee
where $b$ is an axion field, associated with the expansion of the Kalb-Ramond field $B_2$ and $M$ sets the mass scale.  It is this term that manifestly breaks the anti-symmetric combination of discrete symmetry factors, i.e.  $\mathbb{Z}_{Q_-}$,  while the symmetric  combination, i.e. $\mathbb{Z}_{Q_+}$, is preserved.
 
By introducing the redefined fields
\be
A_{1/2} = \frac{1}{\sqrt{2}}\left(A_+ \pm A_- \right)\, , \qquad \alpha_{1/2} = \frac{1}{\sqrt{2}} \left(\alpha_+ \pm \alpha_- \right)
\ee
and for simplicity setting  Higgs fields $\beta_{1/2}=0$ the  sum of  (\ref{covariant}) and (\ref{stuck2})   reads:
%Thus, in the new variables the Lagrangian reads
\be
\mathcal{L} = \frac{1}{2} V_1^2\left(D_\mu \alpha_+ + D_\mu \alpha_- \right)^2 + \frac{1}{2}V_2^2 \left(D_\mu \alpha_+ - D_\mu \alpha_- \right)^2 + M_s^2 \left( D_\mu b \right)^2\, ,  \label{totalL}
\ee
where the covariant derivatives are explicitly given as follows
\be
D_\mu \alpha_\pm := \partial_\mu \alpha_\pm - Q A_\pm\, , \qquad D_\mu b = \partial_\mu b - A_-
\ee
It is evident from (\ref{totalL}) that the effective theory has only {\it one}  discrete gauge symmetry factor $\mathbb{Z}_{Q_+}$, which is a symmetric combination of the two $\mathbb{Z}_{Q}\times \mathbb{Z}_{Q}$ , while the anti-symmetric one is {\it not } present in the effective theory. 

We have therefore demonstrated in the field theory that the Higgsing of a Heterotic  model with two symmetric U(1)'s leads to a Heterotic model with two symmetric  $\mathbb{Z}_Q$, however due to the St\"uckelberg mechanism only  the  symmetric combination of the two $\mathbb{Z}_Q$ is present in the effective theory. It is this surviving discrete symmetry that is matched  via Heterotic/F-theory duality onto the F-theory discrete symmetry. Note that this field theory Higgsing mechanism on the Heterotic side parallels the one on the F-theory side, where only the massless U(1) is visible and gets broken to $\mathbb{Z}_Q$ via Higgs mechanism.

This provides an independent verification of why the anti-symmetric combination of the discrete symmetry factors is not present in the effective theory. In addition, we present for the concrete examples in section \ref{examples} an additional geometrical argument which has been first given in \cite{Aspinwall:2005qw} and is reviewed in detail in subsection \ref{comparingfieldtheoryandgeometry}. As further elaborated at that place, it explains that the elimination of the anti-symmetric combination of  gauge factors in comparable set-ups is due to orbifold singularities of the Heterotic K3 surface. 

In the case of examples where the  two U(1) factors do not appear in a symmetric way, we expect that 
one particular linear combination of two U(1) gauge bosons  becomes massive, while the orthogonal one is massive. However, determination of the St\"uckelberg mass terms of the form (\ref{stuck}) would require a detailed calculation for specific Heterotic compactification  on elliptically fibered  Calabi-Yau threefold $Z_3$ and the calculation of charges of the matter fields responsible for Higgsing to a model with discrete symmetry. Consequently, in this case the origin of the single discrete gauge symmetry factor would be a  specific linear combination of the  two discrete symmetry factors.

\section{Examples} \label{examples}

In this section, we apply the methods from the previous section to construct two examples which realize the discrete symmetries $\mathbb{Z}_2$ and $\mathbb{Z}_3$ within the Heterotic string. For this purpose, we use mirror pairs of K3 surfaces which are promoted to mirror pairs of Calabi-Yau three-folds. In conclusion, we find that the six-dimensional perturbative gauge group associated to one geometry is given as the commutant of the perturbative gauge group of the dual Calabi-Yau within the group E$_8$ $\times$ E$_8$. As has been pointed out in detail in section \ref{Stuckelbergmechanism}, it is important to stress that under Heterotic/F-theory duality only massless gauge degrees of freedem are matched. Thus, there appears sometimes just the symmetric combination of certain gauge factors.

For the $\mathbb{Z}_2$-model, we start with a K3 surface $X_2$ that has E$_7$ $\times$ E$_7$ $\times$ SU(2) $\times$ SU(2) gauge symmetry. Its mirror $\tilde X_2$ has  gauge symmetry U(1) where the U(1) originates from an additional section. As a next step, we promote the mirror pair $X_2, \tilde X_2$ of K3 surfaces to a mirror pair of three-folds, named $X_3, \tilde X_3$. Here, $\tilde X_3$ exhibits a two-section, which originates from the former two sections in $\tilde X_2$, that have been glued together, and indicates a $\mathbb{Z}_2$-symmetry. In addition, there are two non-perturbative SU(2) factors. 

The construction of the $\mathbb{Z}_3$-model is quite parallel to the first one, but it does not exhibit any non-perturbative factors in constrast to the former one.

\subsection{The model with $\mathbb{Z}_2$ gauge symmetry}
As outlined above, we start with a K3 surface that exhibits $(($E$_7$ $\times$ SU(2)$)/\mathbb{Z}_2)^2$ gauge symmetry and use its mirror dual to construct a Heterotic compactification with discrete gauge symmetry $\mathbb{Z}_2$. Thus, as a first step, we construct a pair of dual polytopes $(\Delta^\circ, \Delta)$. Here, $\Delta^\circ$ is the ambient space of the K3 surface with gauge symmetry $(($E$_7$ $\times$ SU(2)$)/\mathbb{Z}_2)^2$.  

\subsubsection{The geometry with gauge symmetry $(($E$_7$ $\times$ SU(2)$)/\mathbb{Z}_2)^2$}
\begin{figure}
\begin{center}
\includegraphics[width=5cm]{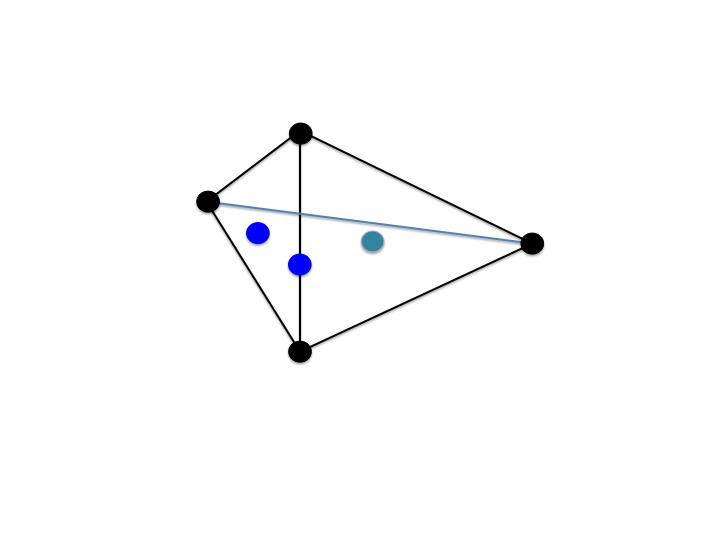}
\includegraphics[width=5cm]{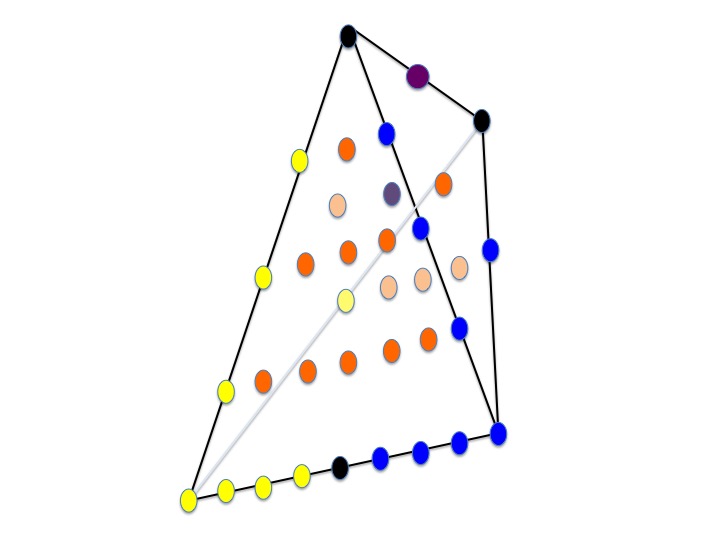}
\end{center}
\caption{The polytope on the left shows the ambient space whose associated hypersurface leads to the $\mathbb{Z}_2$-geometry. The polytope on the right provides the ambient space of the geometry  with gauge symmetry $(($E$_7$ $\times$ SU(2)$)/\mathbb{Z}_2)^2$. The zero plane along which the symplectic cut is performed is marked by the black points. The yellow and blue points give the affine Dynkin diagram of E$_7$. The latter are inherited by the half K3 surfaces $X_2^\pm$, respectively. The purple point corresponds to an SU(2) gauge group which appears in both half K3 surfaces $X_2^\pm$ after the stable degeneration limit. Orange points mark inner points of the facets. Finally, beige-coloured points are on the invisible facets of the polytope.}
\label{PolytopE7E7SU2}
\end{figure}

A very familiar geometry to this one has already been studied in \cite{Cvetic:2015uwu} and we refer to that discussion for some of the details.
Its polytope $\Delta^\circ$ is given by the convex hull of the points
\be
(-2,1,0), \quad (0,1,0), \quad (2,-3,4), \quad (2,-3,-4)\, ,
\ee
while the dual polytope $\Delta$, leading to the geometry with a bi-section is specified by the points 
\be
(-2,-1,0), \quad (0,-1,0), \quad  (1,1,0),\quad (0,-1,-1), \quad (0,-1,1)\, . \label{pointsofDelta}
\ee 
and is shown in figure \ref{PolytopE7E7SU2}. As a first step, we assign the following coordinates to the points of $\Delta^\circ$
\begin{eqnarray}
(-2,1,0) \mapsto y_1, \quad (2,-3,0) \mapsto y_2 \quad (0,1,0) \mapsto y_3, \nn \\  (2,-3,1)\mapsto \tilde U,\quad (2,-3,-1)  \mapsto \tilde V  \, .
\end{eqnarray}
Here $\tilde U, \tilde V$ parameterize the base $\mathbb{P}^1$ of the fibration, while $y_1,y_2,y_3$ are coordinates of the fiber ambient space $\mathbb{P}^{(1,1,2)}/\mathbb{Z}_2$.

Using equation \eqref{CYformula}, we find  the following defining equation
\be
\chi_{\Delta^\circ}: a_1 y_1^4 + a_2 y_1^2 y_2^2 + a_3 y_2^4+a_4 y_3^2 + a_5 y_1 y_2 y_3= 0\, . \label{E7E7SU(2)}
\ee
Here, the coefficients read in terms of $\tilde U, \tilde V$ as follows
\begin{eqnarray}
a_1 &=& a_{11}\, , \nn \\
a_2 &=& a_{21} \tilde U^2 \tilde V^2\, , \nn \\
a_3 &=& a_{31} \tilde U^5 \tilde V^3 + a_{32} \tilde U^4 \tilde V^4 + a_{33} \tilde U^3 \tilde V^5\, , \nn \\
a_4 &=& a_{41} \, , \nn \\
a_5 &=& a_{51} \tilde U \tilde V\, .
\end{eqnarray}
One observes that  there are two sections located at $y_2=0$ and are given by
\be
a_{11} y_1^4 +a_{41} y_3^2 = 0 \, .
\ee
Thus, $\chi_{\Delta^\circ}$ is an elliptically fibered K3. 

In fact, a similar K3 surface has already been investigated in \cite{Cvetic:2015uwu}. To make contact with that description, we transform the constraint \eqref{E7E7SU(2)} into a hypersurface within the ambient space $\mathbb{P}^1 \times \text{Bl}_1 \mathbb{P}^{(1,1,2)}$. To be concrete, the coordinate transformation 
\begin{equation}
y_1 \mapsto \left( x_1^3 x_2 U V \right)^{\frac{1}{4}}, \quad y_2 \mapsto \left( x_1 x_2^3 U^{-3} V^{-3} \right)^{\frac{1}{4}}, \quad y_3 \mapsto x_3 (U V)^{\frac{1}{2}}, \quad (\tilde U, \tilde V) \mapsto (U,V)\, .
\end{equation}
maps \eqref{E7E7SU(2)} onto 
\be
s_1 x_1^3 x_2 + s_2 x_1^2 x_2^2 + s_3 x_1 x_2^3 + s_4x_1 x_2 x_3 + s_5 x_3^2 =0 \, .
\ee
Here, one has
\be
s_1 = a_{11} U V, \quad s_2 = a_{21} U V, \quad s_3= a_{31} U^2 + a_{32} U V + a_{33} V^2, \quad s_4 = a_{51} U V, \quad s_5 = a_{41} U V\, . \label{E7E7SU(2)P112}
\ee
This geometry also has two sections given by $[x_1:x_2:x_3]= [0:1:0]$ and $[x_1:x_2:x_3]= [1:0:0]$. After a transformation into Weierstrass normal form, it becomes transparent that these two sections map onto the zero section and a torsional section of order two, respectively. From the discriminant of the Weierstrass normal form \eqref{Deltaquartic}, one also easily reads off that there are two E$_7$ as well as two SU(2) singularities. Thus, the full gauge group is given by $(($E$_7$ $\times$ SU(2)$)/\mathbb{Z}_2)^2$. 

Note, that there is a second way to read off the gauge group \cite{Candelas:1996su}. The affine Dynkin diagrams of the gauge factors appear in the polytop $\Delta^\circ$ as highlighted in figure \ref{PolytopE7E7SU2}. Note that there is only the non-affine Dynkin diagram visible for the SU(2) factor, as the additional affine node corresponds to a non-toric deformation. It should also be stressed that the construction relies on the fact, that we have admitted a  fiber ambient space different from $\mathbb{P}^{(1,2,3)}$ given by $\mathbb{P}^{(1,1,2)}/ \mathbb{Z}_2$. 

As described in detail in \cite{Cvetic:2015uwu}, the stable degeneration limit decomposes \eqref{E7E7SU(2)P112} into two half K3 surfaces $X_2^+, X_2^-$. Both, $X_2^+, X_2^-$,  inherit an SU(2) $\times$ E$_7$ singularity, as well as the torsional section of order two. 

These findings can be checked by the computation of $h^{(1,1)}(X_{\Delta^\circ})$, which is found to be
\be
h^{(1,1)}(X_{\Delta^\circ}) = 35 -3-1-14+1 = 18\, .
\ee
This accounts for the class of the fiber, the one of the base as well as sixteen resolutional divisors of the corresponding gauge group. This, again, confirms the result that the additional section is torsional. 

In summary, one expects that the dual geometry should give rise to a $\mathbb{Z}_2$ symmetry. This will be the next step of our analysis. As discrete symmetries are visible only in six dimensions in the F-theory description, one needs to fiber the K3 manifold $X_2$ over another $\mathbb{P}^1$. This will eventually promote the two SU(2) singularities to curve of SU(2) singularities resulting in only one SU(2) factor in six dimensions. As argued in \cite{Aspinwall:2005qw} and further elaborated on in subsection \ref{comparingfieldtheoryandgeometry}, that accounts to an identification of the two SU(2) factors such that only their diagonal combination survives.

\subsubsection{The dual geometry with a fiber ambient space $\mathbb{P}^{(1,1,2)}$} \label{DualZ2geometry}

To continue the study of the geometries, we now turn to the analysis of $\Delta$. Here we assign the following coordinates to the points \eqref{pointsofDelta}
\begin{eqnarray}
(-2,-1,0) \mapsto x_1, \quad (-1,-1,0) \mapsto x_4 \quad (0,-1,0) \mapsto x_2, \nn \\  (1,1,0)\mapsto x_3,\quad (0,-1,-1)  \mapsto U, \quad (0,-1,1)  \mapsto V\, .
\end{eqnarray}
Thus, one obtains the following hypersurface constraint: 
\begin{eqnarray}
\chi \!\!\!&\!\!: =\!\!&\!\!\! s_1 x_1^4x_4^3+ s_{2}x_1^3x_2x_4^2+s_{3}x_1^2x_2^2x_4+s_{4}x_1x_2^3+s_{5}x_2^4 \nn \\ \!\!\!&\!\!\!\!&\!\!\! +s_{6}x_1x_2x_3x_4  +s_{7}x_1^2x_3 + s_8 x_2^2x_3  +s_{9}x_3^2x_4  =0\,.  \label{genericsectionP112}
\end{eqnarray}
Here, the $s_i$ take explicitly the form
\begin{eqnarray}
s_1 &=& s_{11} \nn   \\
s_2 &=& s_{21} U^2 + s_{22} U V  + s_{23} V^2\, , \nn \\
s_3 &=& s_{31} U^4 + s_{32} U^3 V + \, ... \, + s_{35} V^4\, , \nn \\
s_4 &=& s_{41} U^6 + s_{42} U^5 V + \, ... \, + s_{47} V^6\, , \nn \\
s_5 &=& s_{51} U^8 + s_{52} U^7 V + \, ... \, + s_{59} V^8\, , \nn \\
s_6 &=& s_{61} U^2 + s_{62} U V  + s_{63} V^2\, , \nn \\
s_7 &=& s_{71}\, , \nn  \\
s_8 &=& s_{81} U^4 + s_{82} U^3 V + \, ... \, + s_{85} V^4\, , \nn \\
s_9 &=& s_{91}\, .   \label{siP112}
\end{eqnarray}
Here, the $s_{ij}$ are complex numbers.
One notices that the above geometry has two sections which are given by $x_1=1, x_2=0$, leading to the equation
\be 
s_{11} + s_{71}x_3 + s_{91} x_3^2 = 0 \label{8Dsections} \, ,
\ee
which is solvable over $\mathbb{C}$. In other words the two sections are located at
\be
S_{1/2} = \left[\sqrt{2s_{91}}: 0: -s_{71} \pm \sqrt{s_{71}^2-s_{11}s_{91}}  \right] \, .
\ee
In addition, the study of the Jacobian reveals that there are no further gauge symmetries. 
 This is confirmed by the dimension of the Picard lattice. Indeed, an application of formula \eqref{homologydimension} reveals that
\be
h^{(1,1)} = 7-3-1-1+1 = 3 \, ,
\ee
which accounts for the base and the class of the elliptic fiber and the additional section. 

The stable degeneration limit is as in reference \cite{Cvetic:2015uwu}  defined by a symplectic cut along the $(x,y,0)$-plane within the coordinate system defined by \eqref{pointsofDelta}; it gives two polytopes $\Delta^+$ and $\Delta^-$.  These polytopes determine  two toric varieties $(\mathbb{P}_{\Sigma^+}, \mathcal{L}^+)$, $(\mathbb{P}_{\Sigma^-}, \mathcal{L}^-)$ which come together with a choice of a line bundle $\mathcal{L}^{\pm}$. A general zero section of the line bundle  $\mathcal{L}^+$  in $\mathbb{P}_{\Sigma^+}$  defines a rational elliptic surface; similarly,  a general section of the line bundle  $\mathcal{L}^-$  defines a rational elliptic surface in  $\mathbb{P}_{\Sigma^-}$.

 However, it is crucial to note that  the appropriate coordinates in the dual fans  are
  \begin{eqnarray}
(-2,-1,0) \mapsto x_1, \quad (-1,-1,0) \mapsto x_4 \quad (0,-1,0) \mapsto x_2, \nn \\  (1,1,0)\mapsto x_3,\quad (0,0,-1)  \mapsto  \lambda^+, \quad (0,-1,1)  \mapsto V\, .
\end{eqnarray}
for  $\Sigma^+$ and
 \begin{eqnarray}
(-2,-1,0) \mapsto x_1, \quad (-1,-1,0) \mapsto x_4 \quad (0,-1,0) \mapsto x_2, \nn \\  (1,1,0)\mapsto x_3,\quad (0,0,+1)  \mapsto  \lambda^-, \quad (0,-1,-1)  \mapsto U\, .
\end{eqnarray}
for $\Sigma^-$.

 Then, if we denote by $v_j \in N$ the lattice point associated to each $ x_j$,  the general section of  the line bundle   $\mathcal{L}^ \pm $ in $\mathbb{P}_{\Sigma^ \pm}$ are:
 \be
\chi^+= \sum_{j} \prod_{m \in P^+} a_P x_{j}^{\langle v_j, P^+\rangle + 1} V^{\langle(0,-1,1), P^+ \rangle+1}  {(\lambda^-)}^{\langle(0,0,-1), P^+ \rangle} 
\ee
 and
  \be
\chi^+= \sum_{j} \prod_{m \in P^-} a_P x_{j}^{\langle v_j, P^-\rangle + 1} U^{\langle(0,-1,-1), P^- \rangle+1}  {(\lambda^+)}^{\langle(0,0,+1), P^- \rangle} \, .
\ee

In summary, one obtains two constraints for the two rational elliptic surfaces (half K3 surfaces).
\begin{eqnarray}
\chi^+ \!\!\!&\!\!: =\!\!&\!\!\! s_1^+ x_1^4x_4^3+ s_{2}^+x_1^3x_2x_4^2+s_{3}^+x_1^2x_2^2x_4+s_{4}^+x_1x_2^3+s_{5}^+x_1^2x_3x_4^2 \nn \\ \!\!\!&\!\!\!\!&\!\!\! +s_{6}^+x_1x_2x_3x_4  +s_{7}^+x_1^2x_3 + s_8^+ x_2^2x_3  +s_{9}^+x_3^2x_4  =0\, , \nn \\
\chi^- \!\!\!&\!\!: =\!\!&\!\!\! s_1^- x_1^4x_4^3+ s_{2}^- x_1^3x_2x_4^2+s_{3}^-x_1^2x_2^2x_4+s_{4}^-x_1x_2^3+s_{5}^-x_1^2x_3x_4^2 \nn \\ \!\!\!&\!\!\!\!&\!\!\! +s_{6}^-x_1x_2x_3x_4  +s_{7}^-x_1^2x_3 + s_8^- x_2^2x_3  +s_{9}^-x_3^2x_4  =0\, ,
\end{eqnarray}
where
\begin{eqnarray}
s_1^+ &=& s_{11} \, , \nn \\
s_2^+ &=& s_{21} U + s_{22} \lambda^+ \, , \nn \\
s_3^+ &=& s_{31} U^2 + s_{32} U \lambda^+ +  s_{33} {\lambda^+}^2 \, , \nn \\
s_4^+ &=& s_{41} U^3 + s_{42} U^2 \lambda^+ + \, ... \, + s_{44} {\lambda^+}^3 \, , \nn \\
s_5^+ &=& s_{51} U^4 + s_{52} U^3 \lambda^+ + \, ... \, + s_{55} {\lambda^+}^4 \, , \nn \\
s_6^+ &=& s_{61} U + s_{62} \lambda^+\, , \nn \\
s_7^+ &=& s_{71} \, , \nn \\
s_8^+ &=& s_{81} U^2 + s_{82} U \lambda^+ + s_{83} {\lambda^+}^2 \, , \nn \\
s_9^+ &=& s_{91} \, .  \label{siP112plus}
\end{eqnarray}
The coefficients for $\chi^-$ are obtained analogously. 
In particular, one  immediately notices that one obtains the correct number of "layers" (i.e. slices parallel to the $x-y$-plane) that are required to obtain an E$_7$ background bundle according to the results of \cite{Berglund:1998ej}. 

We close this section by noting that our geometry under consideration has two sections which lead to an additional U(1) symmetry. However, from an F-theory point of view, it is expected that discrete symmetries become manifest only in six dimensions. Thus, as a final step of our analysis, we compactify our geometry further to six dimensions.

\subsection{The six-dimensional geometry}

As a next step, we investigate a six-dimensional set-up.  For this purpose, we take the direct product of our ambient space \eqref{pointsofDelta} with another $\mathbb{P}^1$. The vertices of the four-dimensional polyhedron $\Delta_4$ read:
\begin{equation}
(-2, -1,  0,  0)\, , ( 1,  1,  0,  0)\, , ( 0, -1, -1,  0)\, , ( 0, -1,  1,  0)\, , ( 0,  0,  0,  1)\, , ( 0,  0,  0, -1)\, .
\end{equation}
By comparing to the Kreuzer-Skarke list \cite{Kreuzer:2000xy}, one finds that $h^{(1,1)}(X_{\Delta_4})$ is five, where $\chi_{\Delta_4}$ denotes the three-fold given as a section of the anti-canonical bundle of the toric ambient space specified by $\Delta_4$. In particular, $\chi_{\Delta_4}$ takes the same form as \eqref{genericsectionP112} and \eqref{siP112}, where the $s_{ij}$  now depend on the two additional $\mathbb{P}^1$ coordinates $[S:T]$. Thus, 
\be
s_{ij}= s_{ij1}S^2 + s_{ij2} ST + s_{ij3} T^2\, .
\ee
Clearly, the two sections defined by \eqref{8Dsections} are no longer well-defined in six-dimensions and form a bi-section. In addition, one finds two SU(2)-singularities located at $s_{91}=0$ which becomes apparent from the explicit equation for the determinant \eqref{Deltaquartic}. As these extend along the Heterotic K3, these contributions to the gauge group are non-perturbative in nature. Summarizing, one finds that the six-dimensional gauge group $G$ is given by
\be
G = G_{\text{pert}} \times G_{\text{non-pert}} =  \mathbb{Z}_2 \times \left(SU(2) \times SU(2) \right)
\ee
This matches $h^{(1,1)}(X_{\Delta_4})=5$, which corresponds to two classes of $\mathbb{P}^1 \times \mathbb{P}^1$, the class of the bi-section and the two resolutional divisors of the SU(2) singularities. We also note that the location of the two SU(2) factors are precisely the points where the two leaves of the bi-section glue together. 

In conclusion, we obtain a perfect match of the six-dimensional perturbative gauge groups of the mirror dual geometries in terms of commutants within E$_8$ $\times$ E$_8$. For the explanation, why there is however only one $\mathbb{Z}_2$-factor, we refer to the following sub-section.

\subsubsection{Comparing field theory and geometry} \label{comparingfieldtheoryandgeometry}

Finally, we comment on the appearance of only one $\mathbb{Z}_2$-factor in six dimensions although there are two background bundles with structure group (E$_7$ $\times$ SU(2))$/ \mathbb{Z}_2$. Thus, naively, one might expect the appearance of an $\mathbb{Z}_2 \times \mathbb{Z}_2$ gauge symmetry. From a field theory perspective, this puzzle gets re-solved by viewing the discrete symmetry as a higgsed version of Abelian groups as explained in detail in section \ref{HiggsingofU1}. In particular, if one turns off the coefficient $s_1$, the fiber ambient space changes from $\mathbb{P}^{(1,1,2)}$ into Bl$_1\mathbb{P}^{(1,1,2)}$. The latter one gives rise to an additional section, such that there are two U(1)-background bundles which are symmetrically embedded into the two E$_8$-factors. As analyzed in \cite{Cvetic:2015uwu, Aspinwall:2005qw}, upon commutation, these background bundles give rise to two U(1) gauge symmetries. One linear combination of these two U(1)-factors turns out to be massive, while the orthogonal linear combination of U(1)-factors is massless. Only the massless factor is seen on the F-theory side and it is this massless U(1) that is higgsed to a discrete symmetry $\mathbb{Z}_2$. 

This discussion is related to the appearance of only one SU(2)-factor in six dimensions which is in fact the dual geometry of the $\mathbb{Z}_2$ considered above. From an algebraic point of view, the two loci of SU(2) singularities get promoted to a curve of SU(2) singularities in six dimensions. A field theory interpretation of this phenomenen has been given in \cite{Aspinwall:2005qw}. Here, the K3 surface acquires orbifold singularities at the locations where it hits the SU(2) locus. Thus, in order to give an interpretation of centralizing a group action that corresponds to the gauge group E$_7$ $\times$ E$_7$ $\times$ SU(2), one has to include an exchange of the SU(2) subgroups of E$_8$ $\times$ E$_8$ in the monodromy of the two bundles around the orbifold points \cite{Aspinwall:2005qw}. It is expected that a similar argument should also hold true in the $\mathbb{Z}_2$ case considered above.
%\Mnote{Antonella, is that correct in our case too? I have some recollection that we do not have a III curve and that there is some sublety about the argument.}
%in contrast to two $\mathbb{Z}_2$-factors that one might naively expect from the commutation of two E$_7$ $\times$ SU(2) background bundles within the respective E$_8$-bundles. This phenomenon can be explained by recalling that the $\mathbb{Z}_2$ has an interpretation as a higgsed U(1). In fact, turning off the coefficient $s_1$ or $s_5$ changes the fiber ambient space such that there are two U(1) bundles which are embedded into one E$_8$-bundle each. However, as explained in \cite{}, only their diagonal combination appears to be massless. As demonstrated in section \ref{HiggsingofU1}, it is the latter which is higgsed to the remaining symmetry $\mathbb{Z}_2$. 

\subsection{The model with $\mathbb{Z}_3$ gauge symmetry}
The construction of the example with discrete $\mathbb{Z}_3$ gauge symmetry parallels the example with $\mathbb{Z}_2$ gauge symmetry and we therefore keep the discussion brief. This time we start with a geometry that has gauge symmetry $(($E$_6$ $\times$ SU(3)$)/\mathbb{Z}_3)^2$. 

\subsubsection{The geometry with $(($E$_6$ $\times$ SU(3)$)/\mathbb{Z}_3)^2$ gauge symmetry}
\begin{figure}
\begin{center}
\includegraphics[width=5cm]{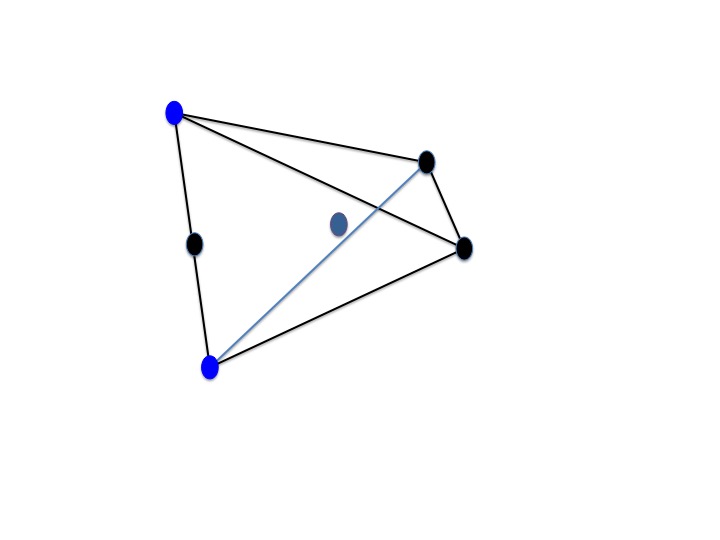}
\includegraphics[width=5cm]{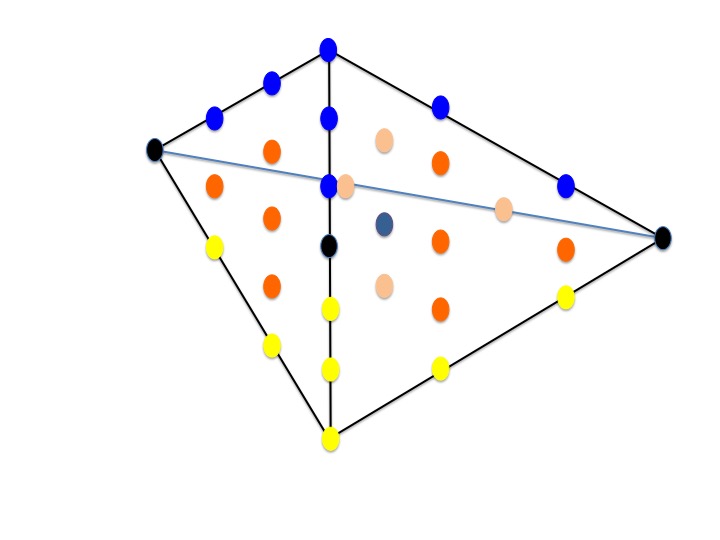}
\end{center}
\caption{The polytope on the left shows the ambient space whose associated hypersurface leads to the $\mathbb{Z}_3$-geometry. The polytope on the right provides the ambient space with gauge symmetry $(($E$_6$ $\times$ SU(3)$)/\mathbb{Z}_3)^2$. The zero plane along which the symplectic cut is performed is marked by the black points.  The yellow and blue points give the affine Dynkin diagram of E$_6$. The latter are inherited by the half K3 surfaces $\chi^\pm$, respectively. Beige-coloured points are on the invisible facets of the polytope. In particular, the two points on the invisible edge correspond to the Dynkin diagram of SU(3) which is inherited by both half K3 surfaces. Finally, orange points mark inner points of the facets and the purple point marks the inner point of the polytope.}
\label{PolytopE6E6SU3}
\end{figure}

We start again with a pair of dual polytopes $(\Delta^\circ, \Delta)$. $\Delta^\circ$ gives rise to a K3 surface with singularity content $(($E$_6$ $\times$ SU(3)$)/\mathbb{Z}_3)^2$, while $\Delta$ gives rise to a K3 surface with fiber ambient space given by $\mathbb{P}^2$. 
$\Delta^\circ$ is given by the convex hull of 
\be
(2,-1,0), \quad (-1,2,0), \quad (-1,-1,3), \quad (-1,-1,-3)\, .\label{pointsofDeltadualZ3}
\ee
while $\Delta$ is given as the convex hull of 
\be
(-1,-1,0), \quad (1,0,0), \quad  (1,1,0),\quad (0,1,0), \quad (-1,-1,1), \quad (-1,-1,-1)\,  . \label{pointsofDeltaZ3}
\ee
The two polytopes are displayed in figure \ref{PolytopE6E6SU3}. 
Assigning coordinates as 
\begin{eqnarray}
(-2,1,0) \mapsto y_1, \quad (2,-3,0) \mapsto y_2 \quad (0,1,0) \mapsto y_3, \nn \\  (2,-3,1)\mapsto \tilde U,\quad (2,-3,-1)  \mapsto \tilde V  \, ,
\end{eqnarray}
an application of formula \eqref{CYformula} reveals that the hypersurface equation for $\chi_{\Delta^\circ}$ is given as
\be
\chi_{\Delta^\circ}: a_1 y_1^4 + a_2 y_1^2 y_2^2 + a_3 y_2^4+a_4 y_3^2 = 0 \, .\label{hypersurfaceE6}
\ee
The coefficients read as follows
\begin{eqnarray}
a_1 &=& a_{11}\, , \nn  \\
a_2 &=& a_{21} \tilde U^2 \tilde V^2\, , \nn  \\
a_3 &=& a_{31} \tilde U^5 \tilde V^3 + a_{32} \tilde U^4 \tilde V^4 + a_{33} \tilde U^3 \tilde V^5\, , \nn  \\
a_4 &=& a_{41} \, .
\end{eqnarray}
The rank of the Picard lattice is found to be $h^{(1,1)}(X_{\Delta^\circ})  = 18$ which accounts for two E$_6$ singularities (to be more precise, its resolutional divisors), two SU(3) singularities, the class of the fiber as well as the base. In addition, the Mordell Weil group equals $\mathbb{Z}_3$. Thus, the full gauge group is given by $(($E$_6$ $\times$ SU(3)$)/\mathbb{Z}_3)^2$. 

Again, after the compactification to six dimensions, the two SU(3) singularities merge into a curve of SU(3) singularities. From the field theory perspective, it is again the symmetric combination of the two SU(3) factors which survives in this limit.

\subsubsection{The dual geometry with fiber ambient space $\mathbb{P}^2$}
We analyse the dual geometry by assigning the following coordinates to the points \eqref{pointsofDelta}
\begin{eqnarray}
(-1,-1,0) \mapsto x_1, \quad (1,0,0) \mapsto x_2 \quad (0,1,0) \mapsto x_3,  \nn  \\  (-1,-1,1)\mapsto U,\quad (-1,-1,-1)  \mapsto V \, .
\end{eqnarray}
In this way, one obtains the following hypersurface constraint: 
\begin{eqnarray}
\chi \!\!\!&\!\!: =\!\!&\!\!\! s_1 x_1^3 + s_2 x_1^2 x_2 + s_3 x_1 x_2^2 + s_4 x_2^3 + s_5 x_1^2 x_3 + s_6 x_1 x_2 x_3 + s_7  x_2^2 x_3 + s_8 x_1 x_3^2 + s_9 x_2 x_3^2 + s_{10} x_3^3 =0\,. \nn \\ &&  \label{genericsectionP2}
\end{eqnarray}
Here, the $s_i$ take explicitly the form
\begin{eqnarray}
s_1 &=& s_{11} U^6 + s_{12} U^5 V + \, ... \, + s_{17} V^6 \, , \nn  \\
s_2 &=& s_{21} U^4 + s_{22} U^3 V + \, ... \, + s_{25} V^4 \, , \nn  \\
s_3 &=& s_{31} U^2 + s_{32} U V  + s_{33} V^2 \, , \nn  \\
s_4 &=& s_{41} \, , \nn  \\
s_5 &=& s_{51} U^4 + s_{52} U^3 V + \, ... \, + s_{55} V^4 \, , \nn  \\
s_6 &=& s_{61} U^2 + s_{62} U V  + s_{63} V^2 \, , \nn  \\
s_7 &=& s_{71} \, , \nn  \\
s_8 &=& s_{81} U^2 + s_{82} U V  + s_{83} V^2 \, , \nn  \\
s_9 &=& s_{91} \, , \nn   \\ \label{siP2}
s_{10} &=& s_{10} \, .
\end{eqnarray}
A closer inspection of this geometry reveals that there are apart from the zero section two further linear independent sections, which is confirmed by the computation of $h^{(1,1)}(X_2)$. In fact, these three sections will glue into a tri-section, once one compactifies further down to six dimensions. We demonstrate that using the Hirzebruch surface $\mathbb{F}_0$ as the base of the fibration. To be more concrete, the vertices of the four-dimensional polyhedron are given by 
\begin{equation}
(-1, -1,  1,  0)\, , ( -1, - 1,  -1,  0)\, , ( 1, 0, 0,  0)\, , ( 0, 1,  0,  0)\, , ( 0,  0,  0,  1)\, , ( 0,  0,  0, -1)\, .
\end{equation}
In contrast to the other example, there are no further non-perturbative enhancements, such that the gauge group is given by $G= \mathbb{Z}_3$. 

\section{Concluding remarks}

In this note we have presented core steps in the understanding of discrete symmetries within the Heterotic/F-theory duality. We propose that for mirror pairs of Calabi-Yau manifolds the gauge group of one geometry is given by the commutant of the gauge group of the dual geometry within E$_8$ $\times$ E$_8$. (However, as explained in detail in section \ref{Stuckelbergmechanism}, only massless gauge fields are matched under Heterotic/F-theory duality, such that sometimes only a symmetric combination of certain gauge factors appears.) Our analysis is based on a two-pronged approach. On the one hand, we have proposed concrete constructions of background bundles whose structure group is given as the commutant of a discrete group within E$_8$. On the other hand, our analysis relies on the field theory investigation of Higgsing of Heterotic compactifications that exhibit U(1) symmetries. For the latter ones, we have restricted ourselves to examples where the U(1) symmetry originates from U(1) background bundles that are symmetrically embedded into both E$_8$ factors. In this way there are two U(1) factors out of which the antisymmetric linear combination is massive due to St\"uckelberg mechanism in six dimensions,  while the symmetric linear combination remains massless. It is the massless U(1) which is spontaneously broken to the discrete symmetry which is to be identified with the F-theory side. 

In addition, we are able to shed  light on the conjecture made in \cite{Klevers:2014bqa, Braun:2014qka, Oehlmann:2016wsb} concerning general F-theory compactifications. It states that mirror symmetry restricted to the elliptic/genus-one fiber exchanges fibrations with multi-sections and geometries that exhibit torsional sections. This is expected from the dual Heterotic side as follows. The commutant of a discrete symmetry $D$ within E$_8$ takes the form $G/D$, where $G$ is a group of ADE type. Using our proposal, which relies on the interpretation of mirror symmetry for K3 surfaces in terms of othogonal lattices \cite{Berglund:1998ej}, we can translate this phenomenon to the exchange of gauge group and structure group on the Heterotic side. In these terms, it is natural that  a multi-section of order $|D|$ gets exchanged with a torsional factor of the Mordell-Weil group of type $D$.

There are many interesting directions this work should be extended to.
It would be desirable to be able to compute charges of matter fields and masses of gauge fields  explicitly in the six-dimensional effective theory on the Heterotic side This would require a deeper understanding of the mathematical structure of the appearing bundles and the computation of their Chern classes. Furthermore,  it would be interesting to prove the mirror symmetry conjecture, which could propably be done building on the results of Belcastro and Dolgachev \cite{Dolgachev, 1998math......9008B}.
%\Mnote{Antonella, how much can we say about Belcastro's thesis here? Can you please add a comment on that?}
\vskip 0.2cm
\noindent  {\bf Acknowledgments}\\
\vskip 0.2cm
{\noindent 
It is a pleasure to thank Peng Song for discussions and collaboration on related topics. 
We also thank Ron Donagi and Denis Klevers for useful discussions. 
We are grateful to the Theory Division of CERN  (M.C. and M.P.),  for  hospitality during the course of the project.  
This research is supported in part by the DOE  Grant Award DE-SC0013528, (M.C., M.P., P.S.), UPenn School of Arts and Sciences Funds  for Faculty Working Group (A.G. and M.C.), the Fay R. and Eugene L.~Langberg Endowed Chair (M.C.) and the Slovenian Research Agency (ARRS) (M.C.).}

\appendix

\section{Weierstrass normal forms}

In this appendix, we summarize the Weierstrass normal form of the most general quartic in $\mathbb{P}^{(1,1,2)}$ as well as that of the most general cubic in $\mathbb{P}^{2}$. We use the following convention for the Weierstrass normal form
\be
y^2 = x^3 + f x + g\, ,
\ee
where the discriminant is given by
\be
\Delta = 4f^3 + 27 g^2\, . \label{DeltaWeierstrass}
\ee
\subsection{The quartic}
The  most general quartic quartic with coordinates $[x_1:x_2:x_3] \in \mathbb{P}^{(1,1,2)}$ is given by
\begin{equation}
\chi:  s_1x_1^4+s_{2}x_1^3x_2+s_{3}x_1^2x_2^2+s_{4}x_1x_2^3 + s_5 x_2^4 +s_{6}x_1^2x_3  +s_{7}x_1x_2x_3  +s_{8}x_2^2x_3  +s_{9}x_3^2  =0\, . 
\end{equation}
The Weierstrass normal form has been determined in, e.g., \cite{Klevers:2014bqa, Huang:2013yta} and is given as
\begin{eqnarray}
f &=& \frac{1}{48} \left(-s_7^4 + 8 s_6 s_7^2 s_8 - 16 s_6^2 s_8^2 + 48 s_5 s_6^2 s_9 - 
   24 s_4 s_6 s_7 s_9 + 8 s_3 s_7^2 s_9 + 16 s_3 s_6 s_8 s_9 \right. \nn \\ && \left. - 24 s_2 s_7 s_8 s_9 + 
   48 s_1 s_8^2 s_9 - 16 s_3^2 s_9^2 + 48 s_2 s_4 s_9^2 - 192 s_1 s_5 s_9^2\right)\, , \nn \\
g &=&  \frac{1}{864} \left(s_7^6 - 12 s_6 s_7^4 s_8 + 48 s_6^2 s_7^2 s_8^2 - 64 s_6^3 s_8^3 - 
   72 s_5 s_6^2 s_7^2 s_9 + 36 s_4 s_6 s_7^3 s_9 - 12 s_3 s_7^4 s_9 \right. \nn \\ && + 
   288 s_5 s_6^3 s_8 s_9 - 144 s_4 s_6^2 s_7 s_8 s_9 + 24 s_3 s_6 s_7^2 s_8 s_9 + 
   36 s_2 s_7^3 s_8 s_9 + 96 s_3 s_6^2 s_8^2 s_9 \nn \\ && - 144 s_2 s_6 s_7 s_8^2 s_9 - 
   72 s_1 s_7^2 s_8^2 s_9 + 288 s_1 s_6 s_8^3 s_9 + 216 s_4^2 s_6^2 s_9^2 - 
   576 s_3 s_5 s_6^2 s_9^2 \nn \\ && - 144 s_3 s_4 s_6 s_7 s_9^2 + 
   864 s_2 s_5 s_6 s_7 s_9^2 + 48 s_3^2 s_7^2 s_9^2 - 72 s_2 s_4 s_7^2 s_9^2 - 
   576 s_1 s_5 s_7^2 s_9^2 \nn \\ && + 96 s_3^2 s_6 s_8 s_9^2 - 144 s_2 s_4 s_6 s_8 s_9^2 - 
   1152 s_1 s_5 s_6 s_8 s_9^2 - 144 s_2 s_3 s_7 s_8 s_9^2 + 
   864 s_1 s_4 s_7 s_8 s_9^2 \nn \\ && + 216 s_2^2 s_8^2 s_9^2 - 576 s_1 s_3 s_8^2 s_9^2 - 
   64 s_3^3 s_9^3 + 288 s_2 s_3 s_4 s_9^3 - 864 s_1 s_4^2 s_9^3 - 
   864 s_2^2 s_5 s_9^3 \nn \\ && \left. + 2304 s_1 s_3 s_5 s_9^3\right) \, ,     \nn \\ 
\Delta &=& -\frac{1}{16} s_9^2 \left(\dots \right)\, . \label{Deltaquartic}
\end{eqnarray}
We note that there is a factor of $s_9^2$ that splits off the remaining polynomial. This is the origin of the non-perturbative SU(2)$^2$ factor as discussed in \ref{DualZ2geometry}. 

\subsection{The cubic}
The  most general quartic quartic with coordinates $[x:y:z] \in \mathbb{P}^{(2)}$ is given by
\begin{equation}
a_1 x^3+ a_2 x^2 y + a_3 x y^2+ a_4 y^3 + a_5 x^2 y + a_6 x y z+ a_7 y^2 z+ a_8 x z^2+ a_9 y z^2+ a_{10} z^3  =0\, . 
\end{equation}
Its Weierstrass normal form reads explicitly
\begin{eqnarray}
f &=& (-a_6^4 + 
   24 a_{10} (2 a_3^2 a_5 - 6 a_2 a_4 a_5 + 9 a_1 a_4 a_6 + 2 a_2^2 a_7 - 
      a_3 (a_2 a_6 + 6 a_1 a_7)) \nn \\ && + 8 a_6^2 (a_5 a_7 + a_3 a_8 + a_2 a_9) - 
   24 a_6 (a_4 a_5 a_8 + a_2 a_7 a_8 + a_3 a_5 a_9 + a_1 a_7 a_9) \nn \\ && + 
   16 (-a_3^2 a_8^2 + 3 a_2 a_4 a_8^2 + a_2 a_3 a_8 a_9 - a_2^2 a_9^2 + 
      a_5 a_7 (a_3 a_8 + a_2 a_9)\nn \\ && - a_5^2 (a_7^2 - 3 a_4 a_9) + 
      3 a_1 (a_7^2 a_8 - 3 a_4 a_8 a_9 + a_3 a_9^2)))\, , \nn \\
g &=&       (216 a_{10}^2 (-a_2^2 a_3^2 + 4 a_2^3 a_4 - 18 a_1 a_2 a_3 a_4 + 
     a_1 (4 a_3^3 + 27 a_1 a_4^2)) - a_6^6 \nn \\ &&+ 
  12 a_6^4 (a_5 a_7 + a_3 a_8 + a_2 a_9) - 
  36 a_6^3 (a_4 a_5 a_8 + a_2 a_7 a_8 + a_3 a_5 a_9 + a_1 a_7 a_9) \nn \\ && - 
  24 a_6^2 (2 a_3^2 a_8^2 - 3 a_2 a_4 a_8^2 + a_2 a_3 a_8 a_9 + 2 a_2^2 a_9^2 + 
     a_5 a_7 (a_3 a_8 + a_2 a_9) + a_5^2 (2 a_7^2 - 3 a_4 a_9) \nn \\ && - 
     3 a_1 (a_7^2 a_8 + 9 a_4 a_8 a_9 + a_3 a_9^2)) + 
  144 a_6 (a_2^2 a_7 a_8 a_9 + (a_1 a_5 a_7^2 + a_3^2 a_5 a_8 \nn \\ && + 
        a_3 a_7 (a_5^2 - 5 a_1 a_8)) a_9 + 
     a_2 (a_5 a_7^2 a_8 + a_3 a_7 a_8^2 + a_3 a_5 a_9^2 + a_1 a_7 a_9^2) \nn \\ && + 
     a_4 (a_5^2 a_7 a_8 - 6 a_1 a_7 a_8^2 + 
        a_5 (a_3 a_8^2 - 5 a_2 a_8 a_9 - 6 a_1 a_9^2))) \nn \\ && + 
  8 (72 a_1 a_3 a_7^2 a_8^2 + 8 a_3^3 a_8^3 + 108 a_1 a_4^2 a_8^3 - 
     108 a_1 a_3 a_4 a_8^2 a_9 - 27 a_1^2 a_7^2 a_9^2 + 72 a_1 a_3^2 a_8 a_9^2 \nn \\ &&  + 
     8 a_2^3 a_9^3 + 108 a_1^2 a_4 a_9^3 + 4 a_5^3 (2 a_7^3 - 9 a_4 a_7 a_9) - 
     3 a_5^2 (4 a_3 a_7^2 a_8 + 9 a_4^2 a_8^2 + 4 a_2 a_7^2 a_9 \nn \\ && - 
        6 a_3 a_4 a_8 a_9 + 9 a_3^2 a_9^2 - 24 a_2 a_4 a_9^2) - 
     3 a_2^2 a_8 (9 a_7^2 a_8 + 4 a_9 (-6 a_4 a_8 + a_3 a_9)) \nn \\ && - 
     6 a_2 (2 a_3^2 a_8^2 a_9 + 3 a_1 a_8 a_9 (-a_7^2 + 6 a_4 a_9) + 
        6 a_3 (a_4 a_8^3 + a_1 a_9^3)) - 
     6 a_5 a_7 (2 a_3^2 a_8^2 \nn \\ && + a_2 a_3 a_8 a_9 + 
        a_2 (-3 a_4 a_8^2 + 2 a_2 a_9^2) + 
        3 a_1 (2 a_7^2 a_8 - a_9 (9 a_4 a_8 + a_3 a_9)))) \nn \\ && + 
  36 a_{10} (-8 a_3^3 a_5 a_8 + 12 a_4^2 (2 a_5^3 - 9 a_1 a_5 a_8) + 
     2 a_7 (-12 a_1 a_2 a_6 a_7 + 12 a_1^2 a_7^2 \nn \\ && + a_2^2 (a_6^2 + 8 a_5 a_7) - 
        4 a_2^3 a_9) + 
     a_3 (6 a_1 a_7 (3 a_6^2 - 4 a_5 a_7) + 4 a_2^2 (a_7 a_8 + a_6 a_9) \nn \\ && - 
        a_2 (a_6^3 + 20 a_5 a_6 a_7 - 36 a_1 a_7 a_9)) - 
     3 a_4 (8 a_2^2 (a_6 a_8 + a_5 a_9) + 
        a_1 (5 a_6^3 - 12 a_5 a_6 a_7 + \nn \\ && 36 a_1 a_7 a_9) + 
        4 a_3 (2 a_5^2 a_6 - 3 a_1 a_6 a_8 - 3 a_5 (a_2 a_8 + a_1 a_9)) + 
        2 a_2 (-3 a_5 a_6^2 + 4 a_5^2 a_7 \nn \\ && - 6 a_1 (a_7 a_8 + a_6 a_9))) + 
     2 a_3^2 (8 a_5^2 a_7 + a_5 (a_6^2 + 2 a_2 a_9) + 
        2 (a_2 a_6 a_8 - 6 a_1 (a_7 a_8 + a_6 a_9))))) \, .  \nn \\    &&
\end{eqnarray}
Using these explicit expressions for $f$ and $g$, one can construct the discriminant \eqref{DeltaWeierstrass}, however, in contrast to the quartic, it does not enjoy any particular factorization properties. 

\bibliographystyle{utphys}	% (uses file "plain.bst")
\bibliography{ref}

\end{document}